\documentclass[a4paper,11pt]{article}
\usepackage{pos}
\usepackage{float}

\title{VERITAS Follow-Up Observations of the
Ultra-High-Energy Neutrino Event KM3-230213A}
\ShortTitle{VERITAS Follow-Up Observations of the UHE Neutrino Event KM3-230213A}

\author*[a]{Connor Mooney for the VERITAS Collaboration}

\affiliation[a]{Department of Physics \& Astronomy, University of Delaware. 104 The Green, Newark, DE 19716, USA.\\}

\emailAdd{comooney@udel.edu}

\abstract{The recent announcement of the detection of the ultra-high-energy (UHE) neutrino event KM3-230213A by the KM3NeT telescope represents a critical opportunity to explore the origins of cosmic neutrinos and their potential gamma-ray counterparts. With an inferred neutrino energy exceeding 100 PeV, this event stands as the most energetic neutrino observed to date. The large offset from the galactic plane (11°) and presence of several blazars with temporally-correlated multi-wavelength counterparts within the 3° localization region raises the possibility of an extragalactic origin. Additionally, the event’s apparent tension with IceCube constraints suggests that it could be transient in nature rather than cosmogenic. VERITAS conducted a targeted follow-up campaign to search for very-high-energy (VHE, >100 GeV) gamma-ray emission associated with KM3-230213A. Observations were performed in February and March 2025, using a four-point wobble strategy centered on the best-fit neutrino position, covering nearly the entire 90\% confidence region. These observations probe potential hadronic gamma-ray emission from a common origin with the neutrino, placing constraints on particle-acceleration scenarios. We present the results of this search, including upper limits on VHE gamma-ray flux and their implications for possible source models of KM3-230213A.}

\ConferenceLogo{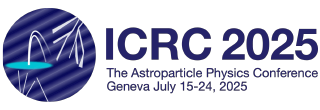}

\FullConference{39th International Cosmic Ray Conference (ICRC2025)\\
 15–24 July 2025\\
Geneva, Switzerland\\}

\begin{document}
\maketitle
\renewcommand{\printHeadAuthors}{Connor Mooney}

\section{Introduction and Motivation}
On February 13, 2025, the KM3NeT Collaboration reported the detection of the ultra-high-energy (UHE) neutrino event KM3-230213A, which had been recorded two years earlier, on February 13, 2023. With a reconstructed muon energy of $120^{+110}_{-60}$ PeV and an inferred median neutrino energy of $220^{+790}_{-110}$ PeV, KM3-230213A is by far the most energetic neutrino event ever reported. The reconstructed arrival direction corresponds to RA = $94.3^\circ$, Dec = $-7.8^\circ$ (J2000), with a 90\% containment radius of $2.2^\circ$. Its extreme energy and the rare, near-horizontal geometry of the track mark KM3-230213A as a benchmark event for studying cosmic neutrino origins, particularly in the poorly explored 100 TeV–PeV regime.

The astrophysical origin of UHE neutrinos remains an open question, with candidate sources ranging from distant active galactic nuclei (AGN) and starburst galaxies to hidden Galactic accelerators and cosmogenic interactions of ultra-high-energy cosmic rays (UHECRs).

High-energy neutrinos are expected to originate in environments where hadronic interactions are efficient, such as shock fronts or magnetized jets. These same interactions produce gamma rays through the decay of neutral pions. If the gamma rays escape the source and are not heavily attenuated by internal absorption or by interactions with the extragalactic background light (EBL), they may be detectable at Earth. This link between neutrinos and gamma rays motivates prompt observations with very-high-energy (VHE; E >100 GeV) gamma-ray telescopes.

This multi-messenger connection was notably demonstrated in 2017 with the detection of the IceCube-170922A neutrino, which was found to be spatially and temporally coincident with a flaring episode from the blazar TXS 0506+056 at 3$\sigma$ \cite{txs_alert}. Multiwavelength observations revealed enhanced emission from the source across X-ray, GeV, and TeV bands, marking the first plausible identification of an astrophysical neutrino source. VERITAS subsequently detected VHE gamma-ray emission from TXS 0506+056 at 5.8$\sigma$ during a follow-up campaign, providing further support for hadronic processes in blazar jets~\cite{txs_veritas}.

Several interpretations of KM3-230213A’s origin have been proposed in the months following its announcement. A dedicated search for blazar counterparts within the KM3NeT 99\% containment region identified seventeen radio-loud blazar candidates, with three showing signs of elevated activity near the event time~\cite{blazar}, making blazars the source class with the most observational support. A Galactic origin is disfavored due to the 11$^\circ$ offset from the Galactic plane and the absence of nearby TeV sources or known PeVatrons \cite{galaxy}. A cosmogenic scenario—wherein the neutrino is produced via interactions of UHECRs with the CMB or EBL—has also been proposed \cite{cosmogenic}. In this case, TeV gamma rays could still be produced but are expected to be strongly attenuated over cosmological distances. However, recent work \cite{clash} shows that such a cosmogenic-only origin is \textit{most} disfavored by the IceCube diffuse flux measurement, with tension measured at 3.0–3.7$\sigma$, while an astrophysical point-source origin is \textit{least} disfavored at 2.0–2.9$\sigma$. These results suggest that no single origin class currently provides a statistically satisfactory explanation, but active blazars remain the most plausible and testable scenario within the VERITAS observable field.

Motivated by this possibility, VERITAS, a ground-based gamma-ray observatory, performed follow-up observations of the KM3-230213A field beginning a few nights after the alert, two years after the event was recorded by KM3-230213A. The observations reported here target the 90\% localization region of KM3-230213A and provide critical constraints on any associated TeV gamma-ray emission.

\section{VERITAS Observations}

VERITAS (Very Energetic Radiation Imaging Telescope Array System) is an array of four 12-meter imaging atmospheric Cherenkov telescopes located at the Fred Lawrence Whipple Observatory (FLWO) in southern Arizona, USA (31$^\circ$40'N, 110$^\circ$57'W, 1.3 km a.s.l.). Each telescope employs a Davies--Cotton optical design, with a segmented reflector composed of 345 hexagonal mirror facets. These mirrors focus Cherenkov light from air showers onto a focal-plane camera containing 499 photomultiplier tubes (PMTs), yielding a field of view of $3.5^\circ$. The array achieves an energy resolution of approximately 15\% at 1~TeV and can detect sources emitting at 1\% of the Crab Nebula flux in under 25 hours of observation.
\cite{performance}.

VERITAS began follow-up observations of KM3-230213A on the night of February 18, 2025 (MJD 60345), five days after the neutrino event was announced, and continued through March 2, 2025 (MJD 60357). Observations were taken at an average zenith angle of $41^{\circ}$ under good weather and hardware conditions. A modified pointing strategy was used, applying a 0.7$^\circ$ offset in each cardinal direction from the best-fit neutrino position (RA = 94.3$^\circ$, Dec = $-7.8^\circ$), chosen to ensure complete coverage of the KM3NeT 90\% containment region, which has a radius of 2.2$^\circ$ \cite{km3 nature}, while maximizing exposure near the best reconstructed position of KM3-230213A (Figure 1). A total of 1057.7 minutes (17.6 hours) of good-quality data remain after standard quality selection cuts and dead-time correction (13\%). 

\begin{figure}[t]
    \centering
    \hspace*{2.1cm}
    \includegraphics[width=0.9\textwidth]{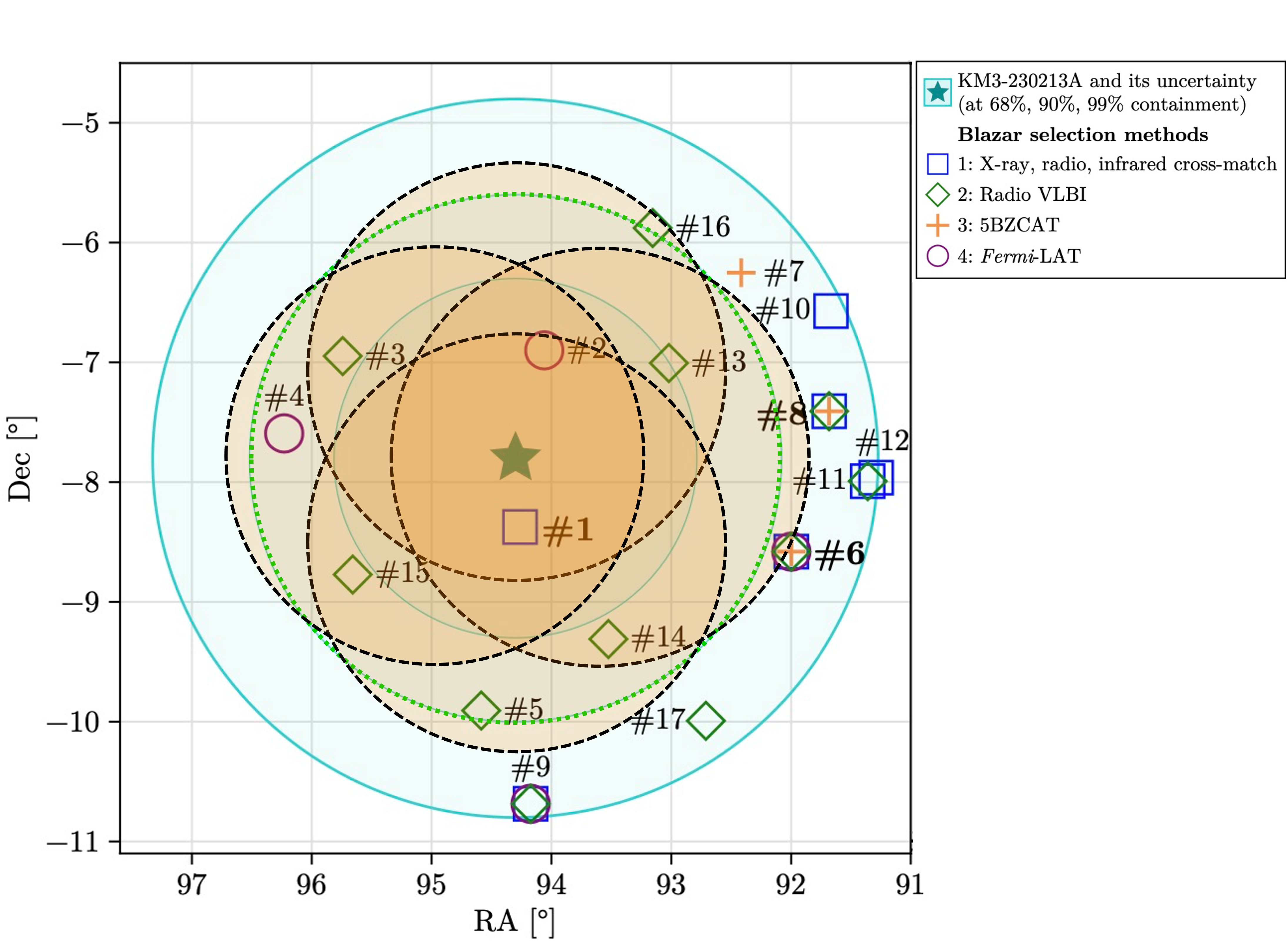}
    \caption{Sky map of KM3-230213A showing the 90\% containment region (green dotted circle), the VERITAS fields of view from observations conducted in wobble mode with 0.7$^\circ$ camera offsets (black dashed circles), and the positions of candidate blazars. Figure adapted from KM3NeT Coll. et al. (2025).}
\end{figure}

The data were processed using the VERITAS EventDisplay analysis software \cite{eventdisplay}, which performs image parameterization, event reconstruction, and gamma/hadron separation using boosted decision trees. The results were cross-validated using a second, independent VERITAS analysis framework, VEGAS \cite{VEGAS}. Gamma/hadron separation cuts were chosen using the standard VERITAS “moderate” configuration, assuming a power-law photon index of $-2.5$. This offers an effective compromise between background rejection and energy threshold, providing good sensitivity over a broad range of source properties.

\section{Results}
Analysis of the 17.6 hours of good-quality data revealed no statistically significant gamma-ray excess anywhere within the $2.2^\circ$ 90\% containment radius of KM3-230213A. The most significant excess in the field reached a cumulative significance of 1.6$\sigma$, which is consistent with expectations from background fluctuations. The significance sky map in Figure 2 shows no hotspots coinciding with known gamma-ray sources or candidate blazars.

\begin{figure}[t]
    \centering
    \includegraphics[width=0.86\textwidth]{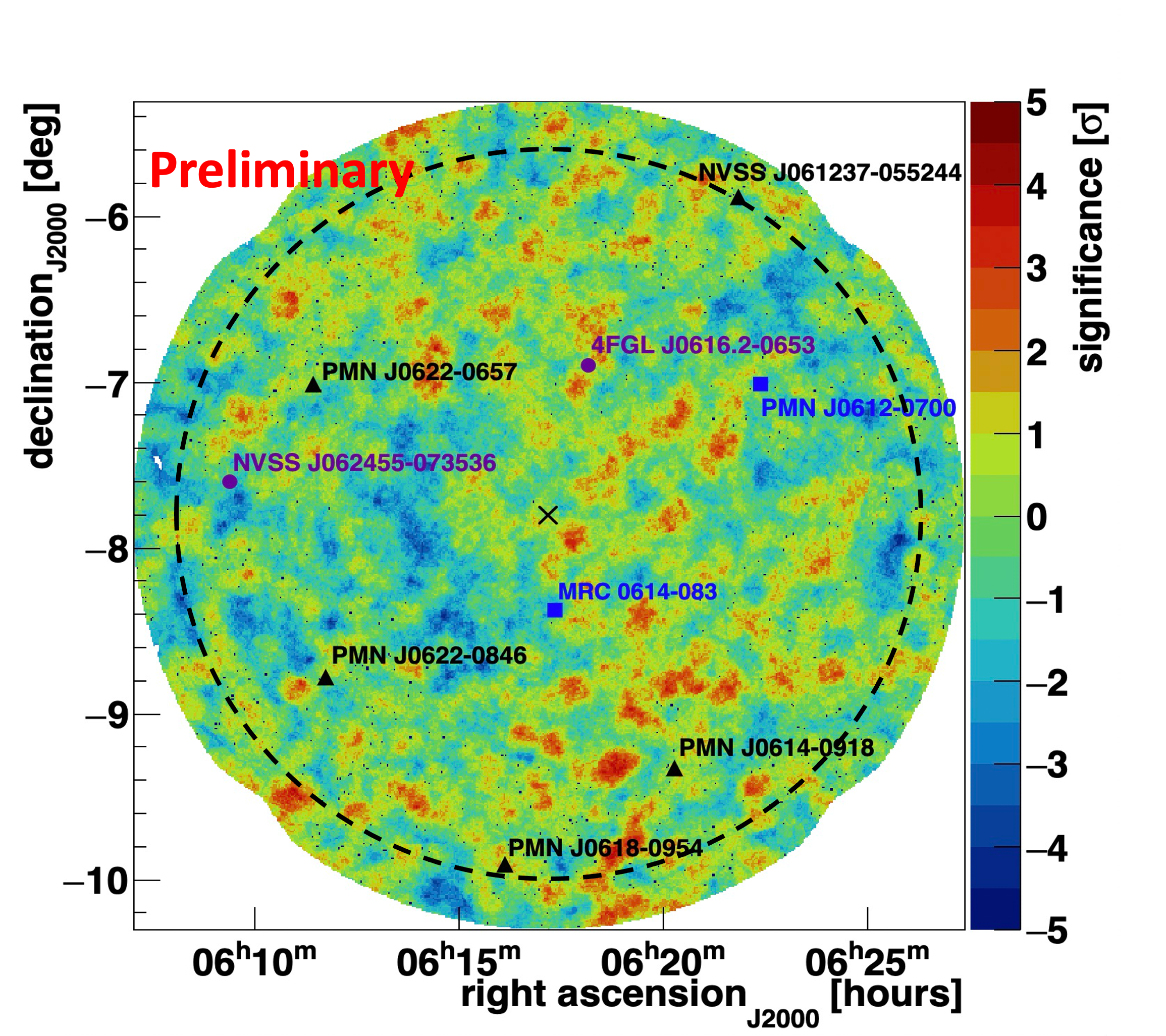}
    \caption{VERITAS significance map of the KM3-230213A field with the 90\% containment region contained in a black dashed circle. Candidate sources are labeled and color-coded by their highest confirmed emission: optical (black triangles), X-ray (blue squares), and GeV gamma-ray (purple circles).}
\end{figure}

To constrain potential gamma-ray emission, a 99\% confidence level flux upper limit was computed at the KM3NeT best-fit coordinates, using the Rolke method \cite{rolke} and assuming a power-law spectrum with photon index $\Gamma = -2.5$. Under the high-zenith observing conditions, the analysis energy threshold was 550 GeV. Significances were also calculated at the positions of candidate blazars proposed in a multiwavelength archival analysis of the KM3-230213A region \cite{blazar}.

At the KM3NeT event location, the signal region contained 62 ON events, while the background region yielded 473 OFF events, scaled by a ON/OFF normalization factor of 0.12. This resulted in a net excess of $4.5 \pm 8.3$ events, corresponding to a significance of 0.5$\sigma$, well below detection threshold. The 99\% confidence upper limit on the integral gamma-ray flux above 550 GeV is $9.49 \times 10^{-13}$\,cm$^{-2}$\,s$^{-1}$, equivalent to 1.81\% of the steady Crab Nebula flux above the same energy.

\section{Discussion and Conclusion}
The absence of significant TeV gamma-ray emission from the direction of KM3-230213A places constraints on potential source classes. Given the sensitivity of VERITAS in the energy range from 200~GeV to 20~TeV, any persistent or temporally coincident TeV-bright source within the 90\% localization region would likely have been detected, provided that the gamma rays could escape both the source environment and intergalactic attenuation. However, short-timescale transients that occurred outside of the VERITAS observation windows cannot be excluded.

The non-detection is consistent with several proposed explanations. For potential blazar counterparts, TeV emission may be suppressed due to internal absorption or attenuation by the extragalactic background light (EBL), particularly if the source lies at high redshift. Alternatively, the source may have been in a low intrinsic emission state during the VERITAS observation window. This was the case for TXS 0506+056, where detection was only possible during a bright flaring episode \cite{txs_veritas}. EBL attenuation is possible for all extragalactic explanations, especially given that none of the 90\% region blazar candidates have a known redshift.

Nine of the seventeen candidate blazars identified within the broader 99\% KM3NeT localization fall inside the 90\% containment region. Among them, MRC 0614-083 stands out as the most plausible counterpart because it lies closest to the best-fit neutrino position and was the only source exhibiting coincident multiwavelength activity, with an eROSITA X-ray flare detected shortly before the neutrino detection \cite{blazar}. However, no significant TeV emission was detected from this source during the observation period, with the measured significance consistent with background fluctuations (see Table 1). The VERITAS observations occurred during a period with no evidence for enhanced X-ray activity, based on sparsely sampled \textit{Swift}-XRT observations shown in Figure 3.

\begin{table}[t!]
    \centering
    \includegraphics[width=\textwidth]{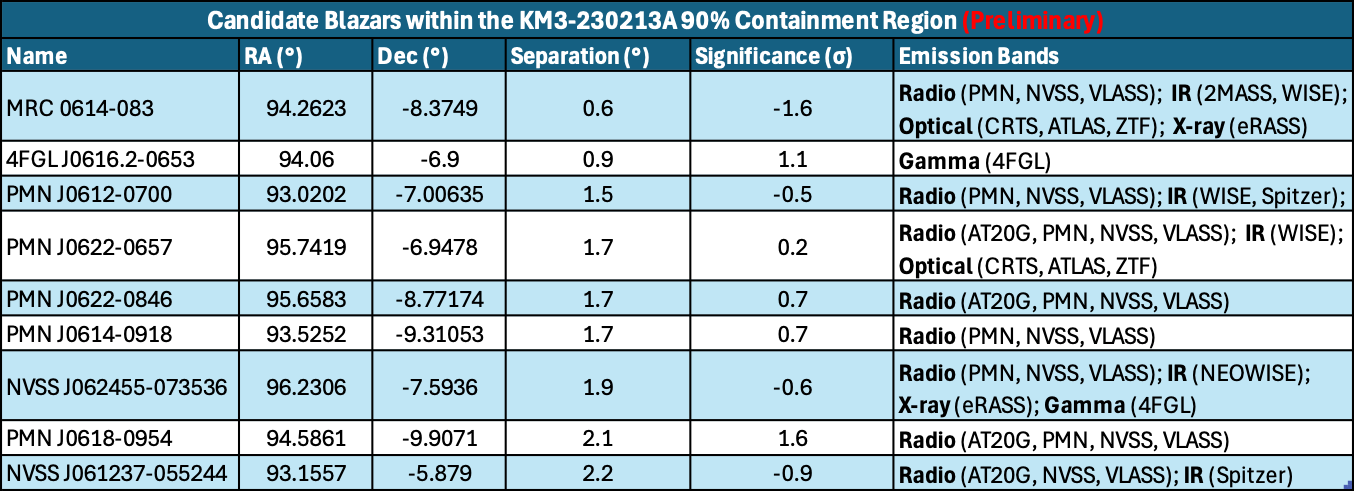}
    \caption{Candidate blazars within the 90\% containment region of KM3-230213A, listing equatorial coordinates, angular separation from the best-fit neutrino position, VERITAS detection significance, and associated multiwavelength counterparts compiled across all available surveys and catalogs.}
\end{table}

\begin{figure}[t!]
    \centering
    \includegraphics[width=0.86\textwidth]{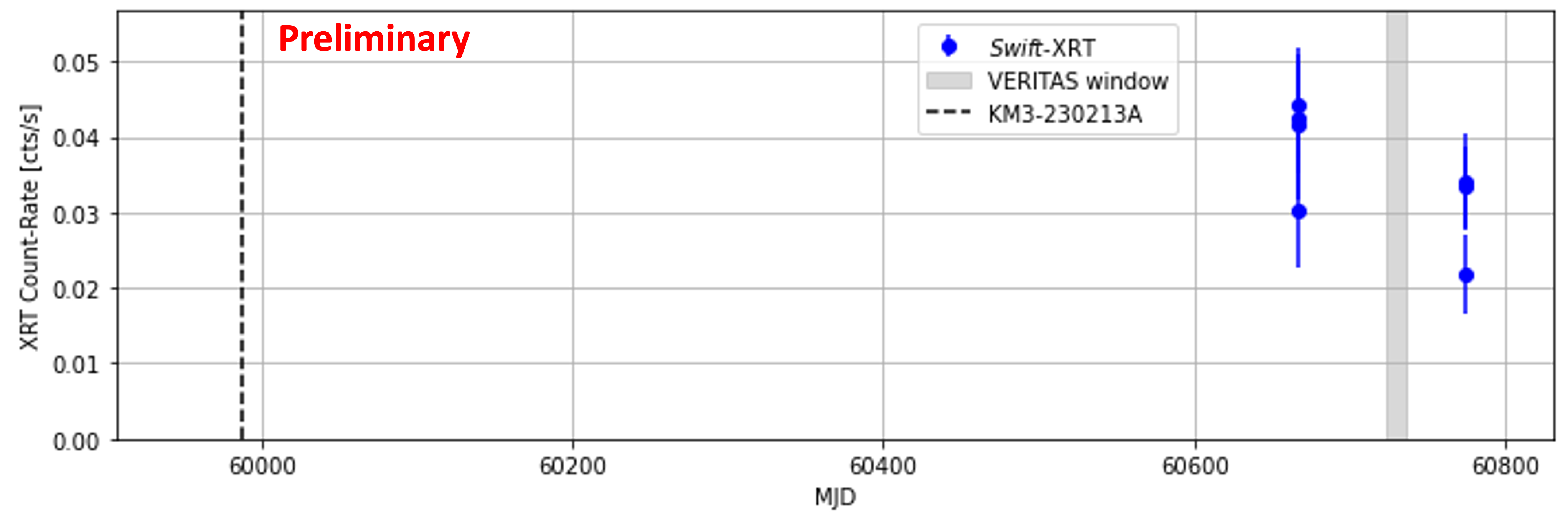}
    \caption{\textit{Swift}-XRT 0.3–10 keV lightcurve of MRC 0614-083. The vertical dashed line marks the time of KM3-230213A. The shaded region shows the VERITAS observation window. Error bars represent 1$\sigma$ statistical uncertainties in count-rate.}
\end{figure}

Note that HAWC also conducted follow-up of KM3-230213A and reported no significant emission in multiple time windows up to one year post-event, placing complementary upper limits at 1 TeV down to $8.06 \times 10^{-13}\,\mathrm{TeV}^{-1}\,\mathrm{cm}^{-2}\,\mathrm{s}^{-1}$ \cite{hawc_atel}. This is consistent with VERITAS' measurement and reinforces the same conclusions.

A transient point source remains one of the least disfavored explanations, as it is most consistent with IceCube’s non-detection at $2\sigma$ tension \cite{clash}. Going forward, improved localization and real-time alerts from KM3NeT, combined with rapid-response coverage from ground-based VHE instruments, will be essential for identifying possible transient sources. Upcoming observatories such as CTAO, and next-generation neutrino facilities like IceCube-Gen2, will further enhance the ability to probe these extreme events with better sensitivity and sky coverage.

\section{Acknowledgements}

This research is supported by grants from the U.S. Department of Energy Office of Science, the
U.S. National Science Foundation and the Smithsonian Institution, by NSERC in Canada, and by
the Helmholtz Association in Germany. This research used resources provided by the Open Science
Grid, which is supported by the National Science Foundation and the U.S. Department of Energy’s
Office of Science, and resources of the National Energy Research Scientific Computing Center
(NERSC), a U.S. Department of Energy Office of Science User Facility operated under Contract
No. DE-AC02-05CH11231. We acknowledge the excellent work of the technical support staff at
the Fred Lawrence Whipple Observatory and at the collaborating institutions in the construction
and operation of the instrument.

\clearpage
\section*{Full Author List: VERITAS Collaboration}

\scriptsize
\noindent
A.~Archer$^{1}$,
P.~Bangale$^{2}$,
J.~T.~Bartkoske$^{3}$,
W.~Benbow$^{4}$,
Y.~Chen$^{5}$,
J.~L.~Christiansen$^{6}$,
A.~J.~Chromey$^{4}$,
A.~Duerr$^{3}$,
M.~Errando$^{7}$,
M.~Escobar~Godoy$^{8}$,
J.~Escudero Pedrosa$^{4}$,
Q.~Feng$^{3}$,
S.~Filbert$^{3}$,
L.~Fortson$^{9}$,
A.~Furniss$^{8}$,
W.~Hanlon$^{4}$,
O.~Hervet$^{8}$,
C.~E.~Hinrichs$^{4,10}$,
J.~Holder$^{11}$,
T.~B.~Humensky$^{12,13}$,
M.~Iskakova$^{7}$,
W.~Jin$^{5}$,
M.~N.~Johnson$^{8}$,
E.~Joshi$^{14}$,
M.~Kertzman$^{1}$,
M.~Kherlakian$^{15}$,
D.~Kieda$^{3}$,
T.~K.~Kleiner$^{14}$,
N.~Korzoun$^{11}$,
S.~Kumar$^{12}$,
M.~J.~Lang$^{16}$,
M.~Lundy$^{17}$,
G.~Maier$^{14}$,
C.~E~McGrath$^{18}$,
P.~Moriarty$^{16}$,
R.~Mukherjee$^{19}$,
W.~Ning$^{5}$,
R.~A.~Ong$^{5}$,
A.~Pandey$^{3}$,
M.~Pohl$^{20,14}$,
E.~Pueschel$^{15}$,
J.~Quinn$^{18}$,
P.~L.~Rabinowitz$^{7}$,
K.~Ragan$^{17}$,
P.~T.~Reynolds$^{21}$,
D.~Ribeiro$^{9}$,
E.~Roache$^{4}$,
I.~Sadeh$^{14}$,
L.~Saha$^{4}$,
H.~Salzmann$^{8}$,
M.~Santander$^{22}$,
G.~H.~Sembroski$^{23}$,
B.~Shen$^{12}$,
M.~Splettstoesser$^{8}$,
A.~K.~Talluri$^{9}$,
S.~Tandon$^{19}$,
J.~V.~Tucci$^{24}$,
J.~Valverde$^{25,13}$,
V.~V.~Vassiliev$^{5}$,
D.~A.~Williams$^{8}$,
S.~L.~Wong$^{17}$,
T.~Yoshikoshi$^{26}$\\
\\
\noindent
$^{1}${Department of Physics and Astronomy, DePauw University, Greencastle, IN 46135-0037, USA}

\noindent
$^{2}${Department of Physics, Temple University, Philadelphia, PA 19122, USA}

\noindent
$^{3}${Department of Physics and Astronomy, University of Utah, Salt Lake City, UT 84112, USA}

\noindent
$^{4}${Center for Astrophysics $|$ Harvard \& Smithsonian, Cambridge, MA 02138, USA}

\noindent
$^{5}${Department of Physics and Astronomy, University of California, Los Angeles, CA 90095, USA}

\noindent
$^{6}${Physics Department, California Polytechnic State University, San Luis Obispo, CA 94307, USA}

\noindent
$^{7}${Department of Physics, Washington University, St. Louis, MO 63130, USA}

\noindent
$^{8}${Santa Cruz Institute for Particle Physics and Department of Physics, University of California, Santa Cruz, CA 95064, USA}

\noindent
$^{9}${School of Physics and Astronomy, University of Minnesota, Minneapolis, MN 55455, USA}

\noindent
$^{10}${Department of Physics and Astronomy, Dartmouth College, 6127 Wilder Laboratory, Hanover, NH 03755 USA}

\noindent
$^{11}${Department of Physics and Astronomy and the Bartol Research Institute, University of Delaware, Newark, DE 19716, USA}

\noindent
$^{12}${Department of Physics, University of Maryland, College Park, MD, USA }

\noindent
$^{13}${NASA GSFC, Greenbelt, MD 20771, USA}

\noindent
$^{14}${DESY, Platanenallee 6, 15738 Zeuthen, Germany}

\noindent
$^{15}${Fakult\"at f\"ur Physik \& Astronomie, Ruhr-Universit\"at Bochum, D-44780 Bochum, Germany}

\noindent
$^{16}${School of Natural Sciences, University of Galway, University Road, Galway, H91 TK33, Ireland}

\noindent
$^{17}${Physics Department, McGill University, Montreal, QC H3A 2T8, Canada}

\noindent
$^{18}${School of Physics, University College Dublin, Belfield, Dublin 4, Ireland}

\noindent
$^{19}${Department of Physics and Astronomy, Barnard College, Columbia University, NY 10027, USA}

\noindent
$^{20}${Institute of Physics and Astronomy, University of Potsdam, 14476 Potsdam-Golm, Germany}

\noindent
$^{21}${Department of Physical Sciences, Munster Technological University, Bishopstown, Cork, T12 P928, Ireland}

\noindent
$^{22}${Department of Physics and Astronomy, University of Alabama, Tuscaloosa, AL 35487, USA}

\noindent
$^{23}${Department of Physics and Astronomy, Purdue University, West Lafayette, IN 47907, USA}

\noindent
$^{24}${Department of Physics, Indiana University Indianapolis, Indianapolis, Indiana 46202, USA}

\noindent
$^{25}${Department of Physics, University of Maryland, Baltimore County, Baltimore MD 21250, USA}

\noindent
$^{26}${Institute for Cosmic Ray Research, University of Tokyo, 5-1-5, Kashiwa-no-ha, Kashiwa, Chiba 277-8582, Japan}

\end{document}